\documentclass[journal]{IEEEtran}
\usepackage{upgreek}

\ifCLASSINFOpdf
  \usepackage[pdftex]{graphicx}
  \graphicspath{{.}}
  \DeclareGraphicsExtensions{.pdf,.jpeg,.png}
\else
  \usepackage[dvips]{graphicx}
  \graphicspath{{../eps/}}
  \DeclareGraphicsExtensions{.eps}
\fi
\newcommand{\unit}[2]{$#1\,\mathrm{#2}$}

\begin{document}
%
\title{Precision Muon Tracking at Future Hadron Colliders with sMDT Chambers}
%
%
%

\author{Oliver Kortner, Hubert Kroha, Felix M\"uller, Sebastian Nowak, Robert Richter\\ \textit{Max-Planck-Institut f\"ur Physik, Munich}}


%
%

\pagenumbering{gobble} 
%



\maketitle

\begin{abstract}
Small-diameter muon drift tube (sMDT) chambers are a cost-effective technology for high-precision muon tracking.
The rate capability of the sMDT chambers has been extensively tested at the Gamma Irradiation Facility at CERN in view of expected rates at future high-energy hadron colliders. 
Results show that it fulfills the requirements over most of the acceptance of muon detectors. 
The optimization of the read-out electronics to further increase the rate capability of the detectors is discussed. 
Chambers of this type are under construction for upgrades of the muon spectrometer of the ATLAS detector at high LHC luminosities. 
Design and construction procedures have been optimized for mass production while providing a precision of better than 10 micrometers in the sense wire positions and the mechanical stability required to cover large areas.  

\end{abstract}


%
\IEEEpeerreviewmaketitle

\section{Introduction}
\begin{figure}[tb]
\centering
\includegraphics[width=0.9\columnwidth]{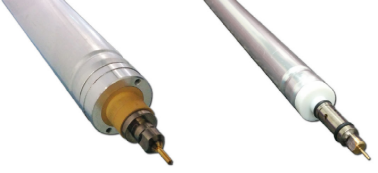}
\caption{Picture of MDT (left) and sMDT (right) tubes.}
\label{fig:tubes}
\end{figure}
Drift tubes are a cost-effective technology for high-precision muon tracking over large detector areas. Monitored drift tube (MDT) chambers have been used for precision muon tracking at the ATLAS experiment. They provide an excellent spatial resolution, high tracking efficiency independent of the track incident angle, and are not susceptible to aging effects from irradition.
Small-diameter muon drift tube (sMDT) chambers~\cite{cite:sMDT} with a tube diameter of \unit{15}{mm}, i.e. half of the tube diameter of the MDT chambers, have been developed. Due to their shorter drift time, they are particularly suitable for high rates in harsh radition environments. A picture of MDT and sMDT is shown in Figure~\ref{fig:tubes}, and a detailed destription of its physics properties will be given in Section~\ref{sec:sMDT}. 

Two future hadron-colliders are considered as application sMDT chambers in this study: the upgrade of the large hadron collider~(LHC) for high luminosities~(HL-LHC), and the future circular hadron collider (FCC-hh). 

The HL-LHC envisages to fully exploit the potential of the LHC. While the design centre-of-mass energy of \unit{14}{TeV} is sustained, the luminosity will be increased from $300\,\mathrm{fb}^{-1}$ at the LHC to $3000\,\mathrm{fb}^{-1}$ at the HL-LHC, which will allow to study very rare processes at high precision. The large accumulated data set is based on a peak instantaneous luminosity of $L=7\cdot10^{34}\,\mathrm{cm}^{-2}\mathrm{s}^{-1}$.
The ATLAS sMDT chambers are part of the upgrade program of the ATLAS detector and will be employed at yet uncovered regions of very high rate. The required momentum resolution for a muon with a transverse momentum of $p_\mathrm{T} = 1\,\mathrm{TeV}$ remains the same as for the LHC, $dp_\mathrm{T} / p_\mathrm{T} = 10\%$.

\begin{figure}[tb]
\centering
\includegraphics[width=0.9\columnwidth]{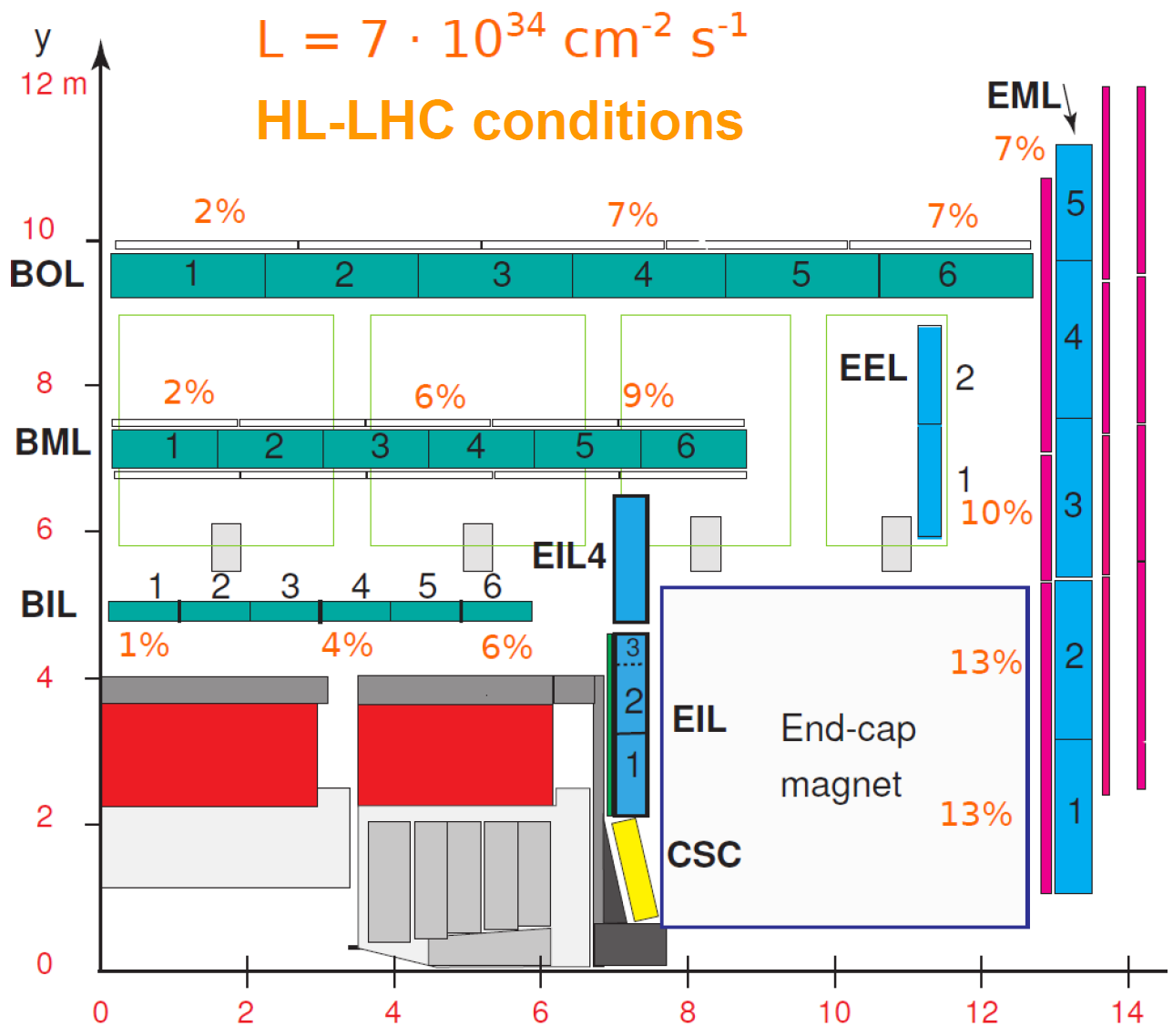}
\caption{Expected occupancies of MDT chambers for operating conditions at the HL-LHC in a quadrant of the ATLAS muon spectrometer.}
\label{fig:occupancy_hllhc}
\end{figure}

The FCC-hh is planned as the successor of the HL-LHC to search for physics processes beyond the Standard Model of Particle Physics at a centre-of-mass energy of $\sqrt{s} = 100\,\mathrm{TeV}$. In the conceptual design study, superconducting dipole magnets with a magnetic field strength of $16\,\mathrm{T}$ are considered to be feasible. The resulting circular tunnel requires a circumference of around $100\,\mathrm{km}$ and is planned to be located at CERN in order to use the existing LHC complex as preaccelerator. Its luminosity is forseen to be increased in two steps. Phase~I corresponds to $L=5\cdot10^{34}\,\mathrm{cm}^{-2}\mathrm{s}^{-1}$, i.e. slightly below the one of the HL-LHC, while Phase~II envisages $L=30\cdot10^{34}\,\mathrm{cm}^{-2}\mathrm{s}^{-1}$, i.e. four times higher than the one at the HL-LHC.
The detector concepts for the FCC-hh are very similar to those of ATLAS and CMS, using a muon spectrometer with either a superconducting toroidal or (twin) solenoid magnet system in the barrel region, and dedicated dipole magnets in the forward region. The air-core magnets are supposed to provide a magnetic field strength of~\unit{3}{T}. In order to account for the seven times higher centre-of-mass energy, a momentum resolution of $dp_\mathrm{T} / p_\mathrm{T} = 10\%$ for $p_\mathrm{T} = 7\,\mathrm{TeV}$ is foreseen. Given the higher field strength at the FCC-hh detectors, the spatial resolution is required to be $30-40$ micrometer, similar to the spatial resolution of MDT chambers at ATLAS. 

The operating conditions of a hadron collider are characterized by high $\gamma$~and neutron background rates from interactions of the proton collision products in the detector, and with the beam shielding. Thus, the radiation depends on the structure of minimum bias events, and on the instantaneous luminosity. The inelastic cross-section of proton-proton collisions at \unit{\sqrt{s}=14}{TeV} is about \unit{\sigma_{\mathrm{pp}} = 80}{mb} with an average number of charge particles of $N_\mathrm{ch} = 5.4$, and an average particle momentum of \unit{p = 0.6}{GeV}. At \unit{\sqrt{s}=100}{TeV}, this increases only moderately to about \unit{\sigma_{\mathrm{pp}} = 100}{mb}, $N_\mathrm{ch} = 8$, and \unit{p = 0.8}{GeV}, respectively. Typical rates in the barrel (end-cap) region for a muon detector according to these operating conditions amount to approximately $20\,\mathrm{Hz}/\mathrm{cm}^{2}$ ($100\,\mathrm{Hz}/\mathrm{cm}^{2}$) at the HL-LHC, and $100\,\mathrm{Hz}/\mathrm{cm}^{2}$ ($250\,\mathrm{Hz}/\mathrm{cm}^{2}$), assuming a detector at the FCC-hh with a design similar to ATLAS. Figure~\ref{fig:occupancy_hllhc} shows the occupanciy of MDT chambers in a quadrant of the ATLAS detector in case of the HL-LHC. While the occupancy for MDT in the barrel region is well below a limit of $30\%$, some areas in the inner end-cap (EIL) need to be replaced.

\begin{figure}[tb]
\centering
\includegraphics[width=0.9\columnwidth]{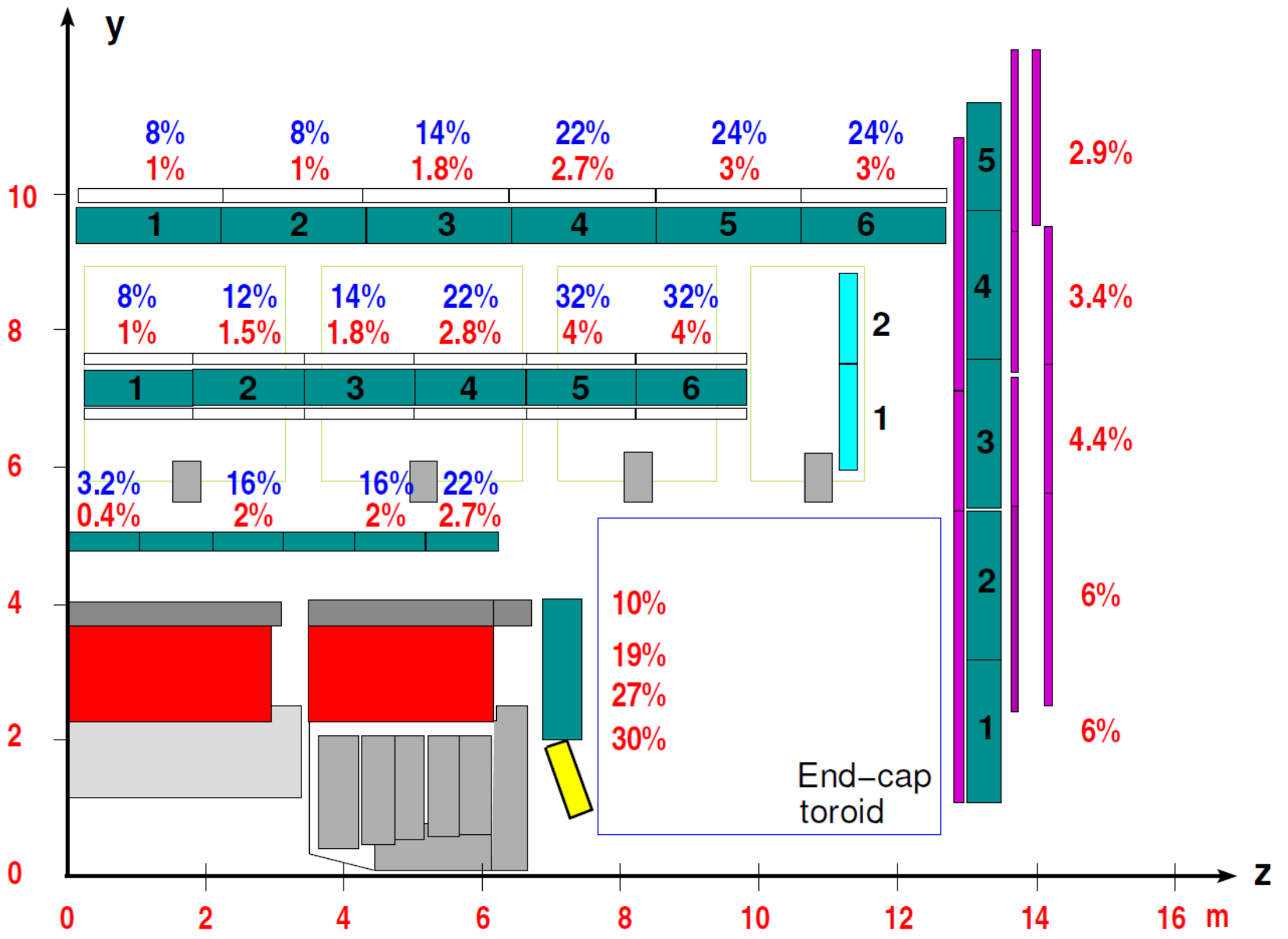}
\caption{Expected occupancies of MDT (blue) and sMDT (red) chambers for operating conditions at the FCC-hh Phase~II luminosity, on the basis of the ATLAS muon spectrometer for illustration.}
\label{fig:occupancy_fcc}
\end{figure}
The smaller drift radius of the sMDT result in higher rate capabilities and improved resolution. Figure~\ref{fig:occupancy_fcc} shows the occupancy in FCC-hh phase~II for MDT and sMDT chambers, assuming a detector design that is identical to the one of ATLAS for illustative purposes. In the barrel, the MDT chamber occupancy ranges from 3.2\% to 32\%, i.e. close to the limit of MDT. The sMDT occupancy, however, stays below 4\% for the barrel, and reaches 30\% only in the inner end-cap where rates are in the order of $250\,\mathrm{Hz}/\mathrm{cm}^{2}$. Hence, sMDT chambers are well suited in terms of expected rates at a future circular collider.

\section{Small-diameter muon drift tubes}
\label{sec:sMDT}
Small-diameter muon drift tubes are made of Al tubes with an outer diameter of \unit{15}{mm} and a sense wire with a diameter of $50\,\upmu\mathrm{m}$. They are operated at a potential of \unit{2700}{V} using a gas mixture of Ar:CO$_{2}$ in a ratio of 93:7 and a pressure of \unit{3}{bar}. The gas gain results in $2 \cdot 10^{4}$. The sMDT represent the further development of the ATLAS MDT, whose key difference is a larger drift tube diameter of~\unit{30}{mm}.
\begin{figure}[tb]
\centering
\includegraphics[width=0.9\columnwidth]{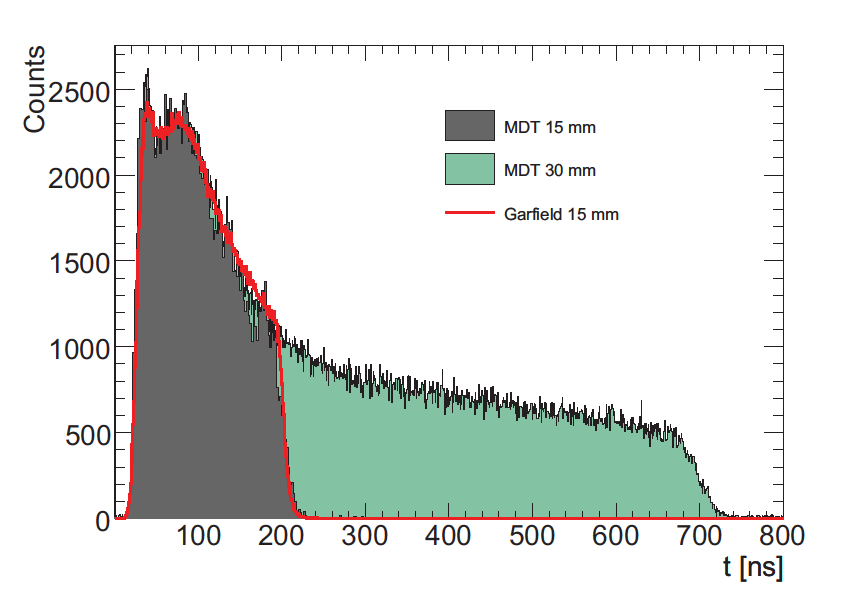}
\caption{Drift time spectrum for MDT (green) and sMDT tubes (grey). Prediction from the GARFIELD simulation~\cite{cite:garfield} for the sMDT tube is shown by the red line~\cite{cite:bittner}.}
\label{fig:spectrum}
\end{figure}
A comparison of the drift time spectra of MDT and sMDT is shown in Figure~\ref{fig:spectrum}. The maximum drift time corresponds to particle tracks close the the tube wall and amounts to \unit{700}{ns} for the MDT. Due to non-linear space-to-drift time relationship $r(t)$, the maximum drift time for the sMDT is reduced to~\unit{200}{ns}. With the smaller area cross-section of the sMDT with respect to the MDT, this improves the drift tube occupancy by a factor of eight.

\section{Drift tube performance at high background rates}
\begin{figure}[tb]
\centering
\includegraphics[width=0.72\columnwidth]{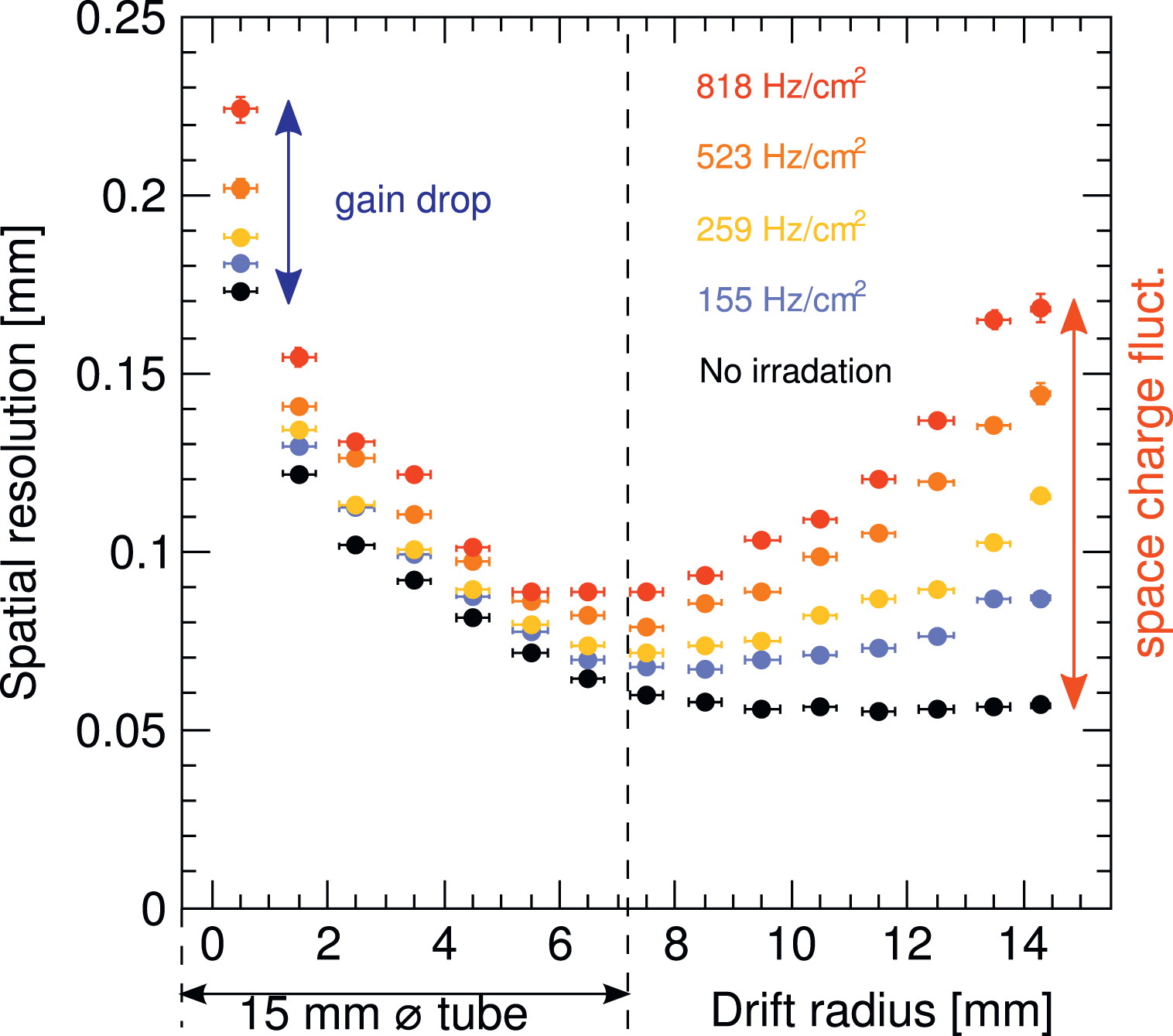}
\caption{Effect of space charges on the spatial resolution for different background irradiation flux~\cite{cite:bittner}.}
\label{fig:spacecharge}
\end{figure}
The drift tube performance at high background rates has been demonstrated in extensive tests~\cite{cite:bittner} at the CERN Gamma Irradiation facility (GIF)~\cite{cite:gif}. No aging up to a charge accumulation on the sense wire of $9\,\mathrm{C}/\mathrm{cm}$ has been found. This corresponds to 15 times the requirement for the current ATLAS MDT chambers~\cite{cite:sMDT}, and also exceeds the requirement for future colliders such as the FCC-hh.

\begin{figure}[tb]
\centering
\includegraphics[width=0.9\columnwidth]{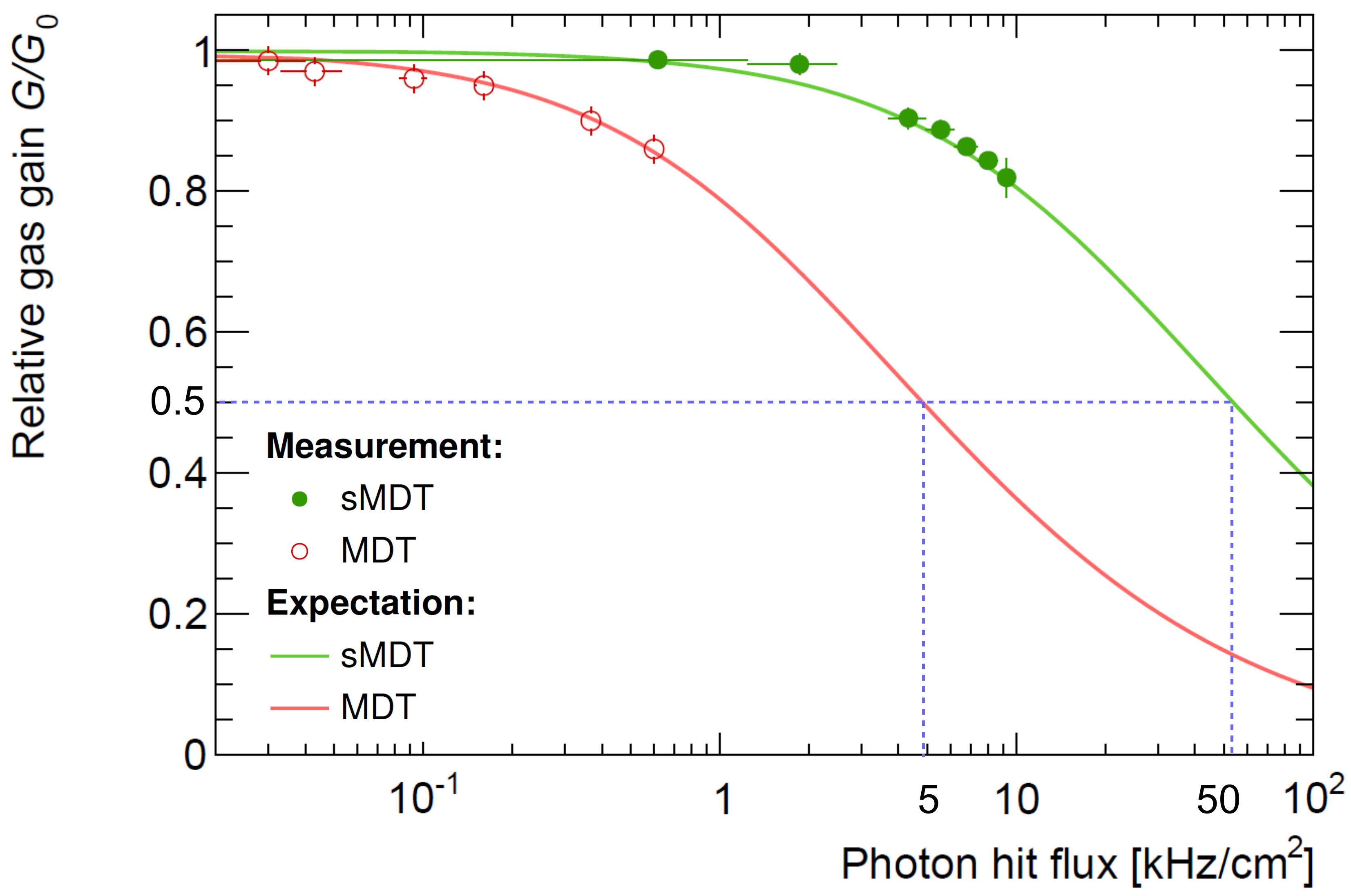}
\caption{Relative gas gain for MDT and sMDT as a function of the photon hit flux with respect to the nominal gas gain $G_0=2\cdot10^4$~\cite{cite:bittner}.}
\label{fig:gaindrop}
\end{figure}

The main effect from the intense irradiation is space charge building up inside the drift tubes. Ions which are created in the avalanche close the anode wire form a constant stream of slowly drifting space charges towards the tube walls. This charge has two effects. The first effect is a shielding of the potential at the anode wire, which reduces the electric field and thus the gas gain. It can be shown that the voltage drop $V$ behaves like $V \propto r^3_\mathrm{max}$, where $r^3_\mathrm{max}$ is the drift tube radius~\cite{cite:schwegler}. The gain drop decreases the signal amplitude and hence the spatial resolution at small drift radii of $r < 5\,\mathrm{mm}$. 
The second effect is due to space charge fluctuations, which modify the local electric field inside the tube, and as a consequence the space-to-drift-time relationship $r(t)$. The degradation of the spatial resolution due to this effect is stronger if the $r(t)$ relationship is non-linear, as it is the case for drift radii $r > 5\,\mathrm{mm}$.
Figure~\ref{fig:spacecharge} shows both effects on the spatial resolution as a function of the drift radius for different irradiation levels from $0$ to $818\,\mathrm{Hz}/\mathrm{cm}^{2}$. A tube diameter of \unit{30}{mm} gives the best resolution at low irradiation, but space charge effects degrade the resolution substantially at small and large drift radii under high irradiation. Small-tube drift tubes with a diameter of \unit{15}{mm} avoid the effects from space charge fluctuations completely, and also the gain drop is largely reduced, which can be seen in Figure~\ref{fig:gaindrop}.  
It shows the gas gain with respect to the nominal gas gain of $G_0=2\cdot10^4$ as a function of the photon flux for the MDT and the sMDT. The expectation is based on Diethorn's formula for a primary ionisation charge of $Q_\mathrm{prim} = 1300e$ ($Q_\mathrm{prim} = 900e$) per converted photon for MDT (sMDT), which has been determined experimentally~\cite{cite:bittner}. For the smaller tube diameter, the gain loss is reduced by an order of magnitude.

\begin{figure}[tb]
\centering
\includegraphics[width=0.9\columnwidth]{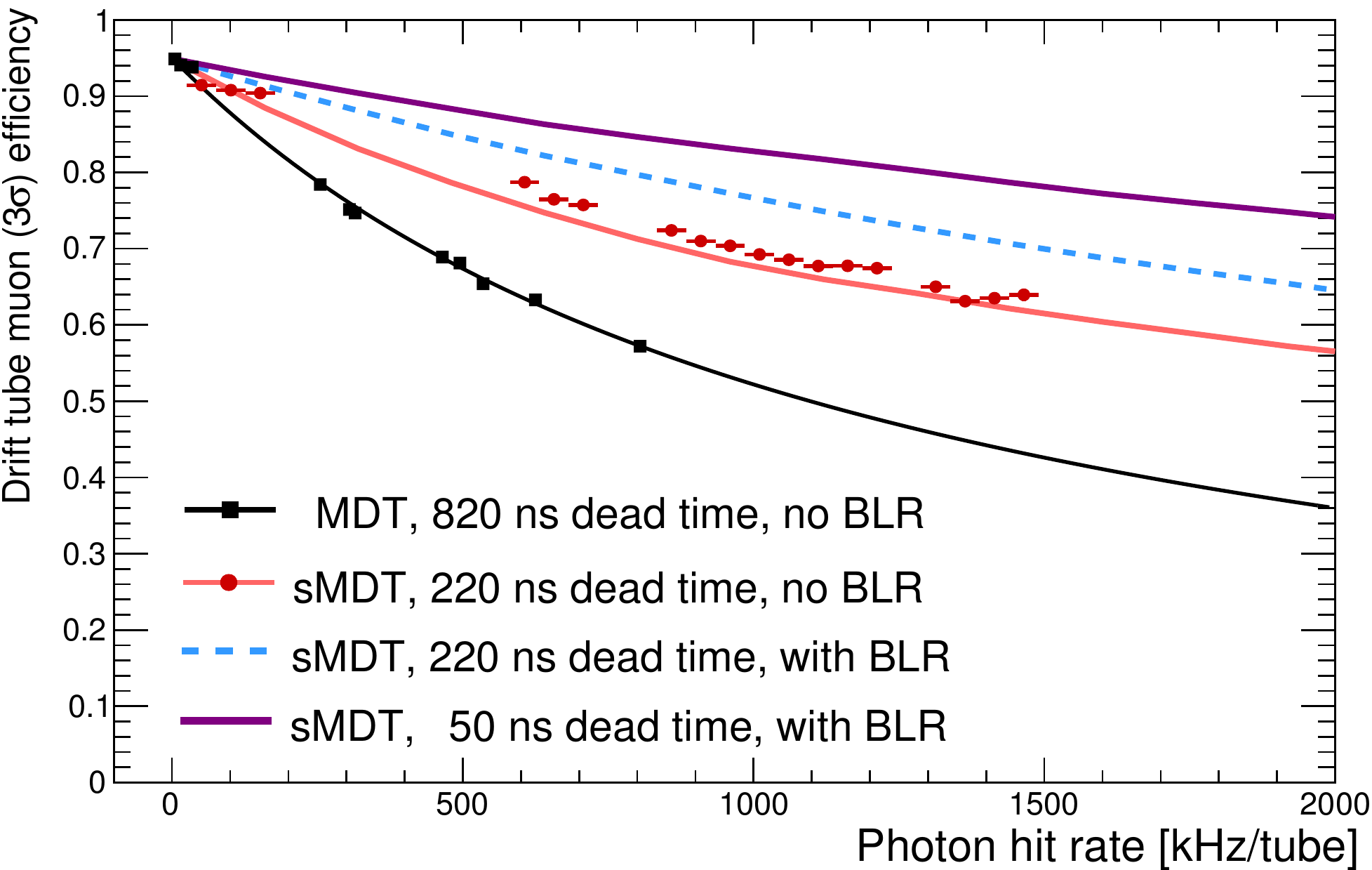}
\caption{Muon detection efficiency of  MDT and sMDT as a function of the photon hit flux per tube. The tube length is \unit{1}{m}. The efficiency defined as the probablity to find a hit on the extrapolated track within $3\sigma$ of the tube spatial resolution. Electronics with and without baseline restoration (BLR) are compared. The curves are from a detailed simulation of detector and electronics response~\cite{cite:bittner}.}
\label{fig:efficiency}
\end{figure}
Figure~\ref{fig:efficiency} shows the muon detection efficiency of MDT and sMDT as a function of the photon hit flux per tube, where the tube length is \unit{1}{m} and the efficiency is defined as the probablity to find a hit on the extrapolated track within $3\sigma$ of the tube spatial resolution. The improvement of sMDT over MDT is noticeable. However, the performance of the sMDT is limited by the read-out electronics, which suffers from signal pile-up effects of consecutive hits. Detailed simulation of detector and electronics response show that pile-up effects can be mitigated with an improved electronics.

\begin{figure}[tb]
\centering
\includegraphics[width=0.9\columnwidth]{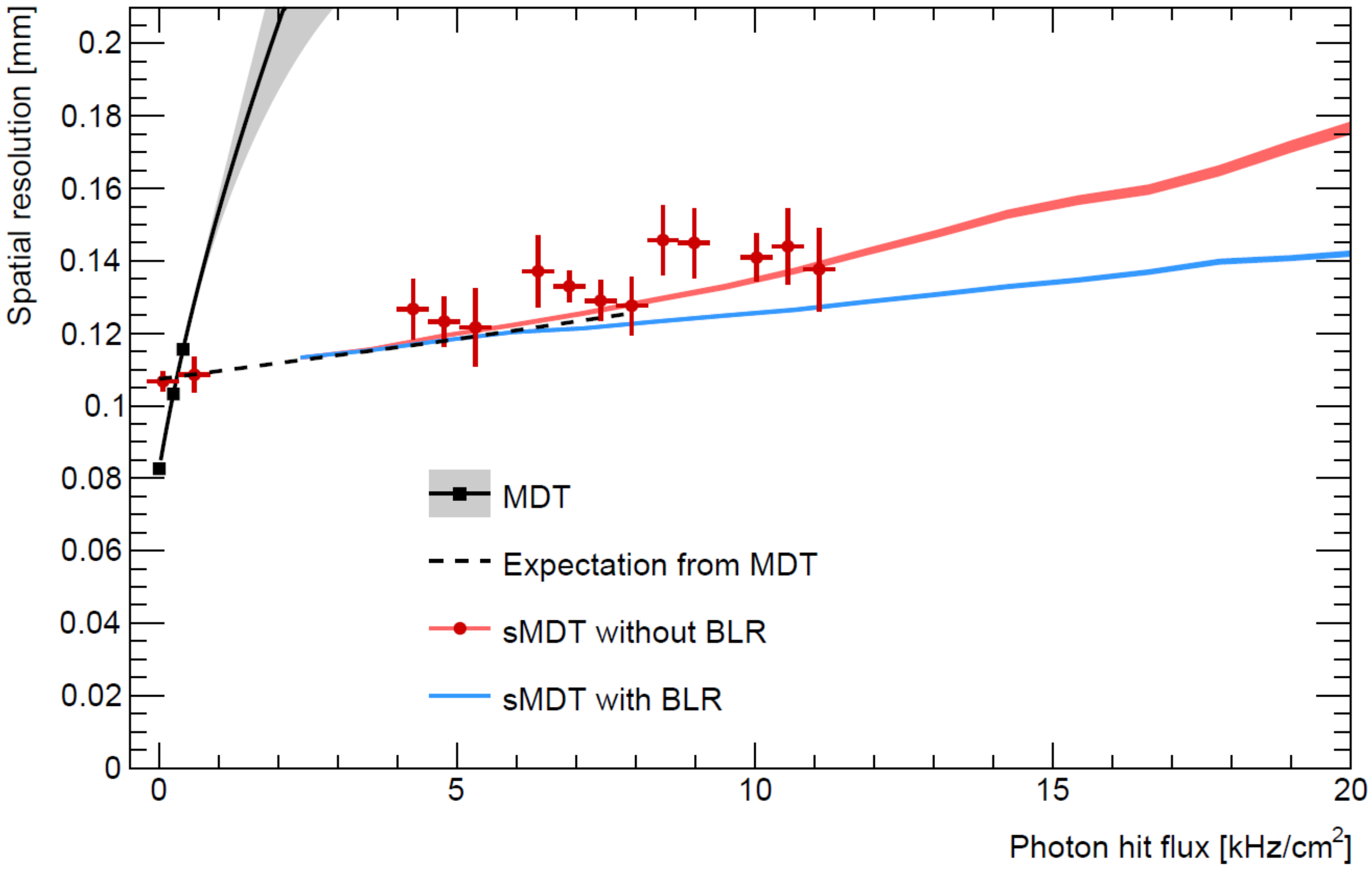}
\caption{Spatial resolution of drift tubes as a function of the background hit rate for MDT and sMDT using signal shaping with and without baseline restoration (BLR). The markers represent the measurements at the GIF, while the lines are produced using a full simulation of the drift tube response including read-out electronics~\cite{cite:bittner}.}
\label{fig:resolution}
\end{figure}
Similarly, Figure~\ref{fig:resolution} shows the advantage of the sMDT against the MDT for the spatial resolution of a single tube as a function of the background hit flux. While the MDT strongly depends on the photon hit flux, the sMDT distribution is almost independent. Further improvements can be achieved using a modified electronics, as shown by a full simulation. 

\section{Fast baseline restoration}
\begin{figure}[tb]
\centering
\includegraphics[width=0.7\columnwidth]{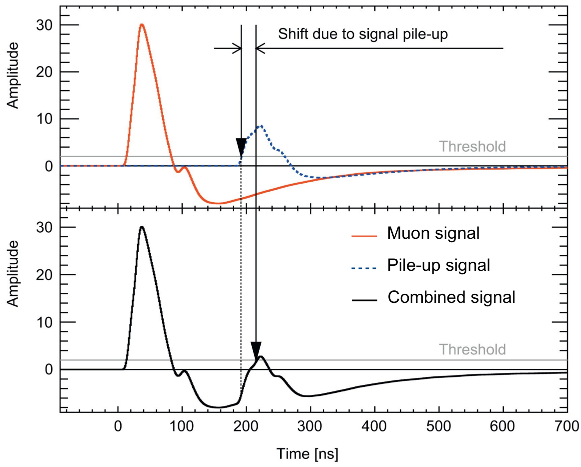}
\caption{Illustration of signal pile-up effects with bipolar shaping  at high counting rates. The individual pulses are shown in read and blue, while the superposition of the two is shown in black. The signal pile-up leads to a reduction of the amplitude of successive hits and a jitter in the drift-time measurement.}
\label{fig:pileup}
\end{figure}
With the much shorter maximum drift time of the sMDT compared to the MDT, the deadtime of the read-out electronics can be reduced from about \unit{700}{ns} for the MDT to \unit{200}{ns} for the sMDT. However, the improvement of the muon detection efficiency and spatial resolution at high counting rates is limited by the existing read-out electronics of the MDT. The 8-channel analog front-end chip for amplification, shaping and discrimination of the signal (ASD) employs a bipolar shaping to guarantee baseline stability. The typical baseline restoration time is in the order of \unit{500}{ns}, well below the drift time of the MDT, but significantly longer than the drift time of the sMDT. For photon hit rates in the order of $10\,\mathrm{kHz}/\mathrm{cm}^{2}$, the average interval between two consequtive hits in a muon tube is about $1\,\upmu\mathrm{s}$, which exhibits a significant probablity for signal pile-up in the interval from \unit{200}{ns} to \unit{500}{ns} after the first hit. Figure~\ref{fig:pileup} illustrates the problem. The signal pile-up during the undershoot leads to a reduction of the amplitude of successive hits and a jitter in the drift-time measurement or even a complete loss of signal if the reduced amplitude is below threshold.

A concept to suppress the signal pile-up during the undershoot is the fast baseline restoration (BLR). 
In a simple circuit of a BLR~\cite{cite:seattle}, the signal is sent through two capacitors $C_1$ and $C_2$, where a diode in reverse direction is connected to ground in between.
The diode current $I_\mathrm{base}$ sets a positive working point between $C_1$ and $C_2$. For positive input signals, the working point is further increased, and no current is drawn in reverse direction of the diode. Hence the signal remains unchanged. Negative input signals, however, shift the working point into the conducting region of the diode, leading to a fast restoration of the signal baseline.

\begin{figure}[tb]
\centering
\includegraphics[width=0.7\columnwidth]{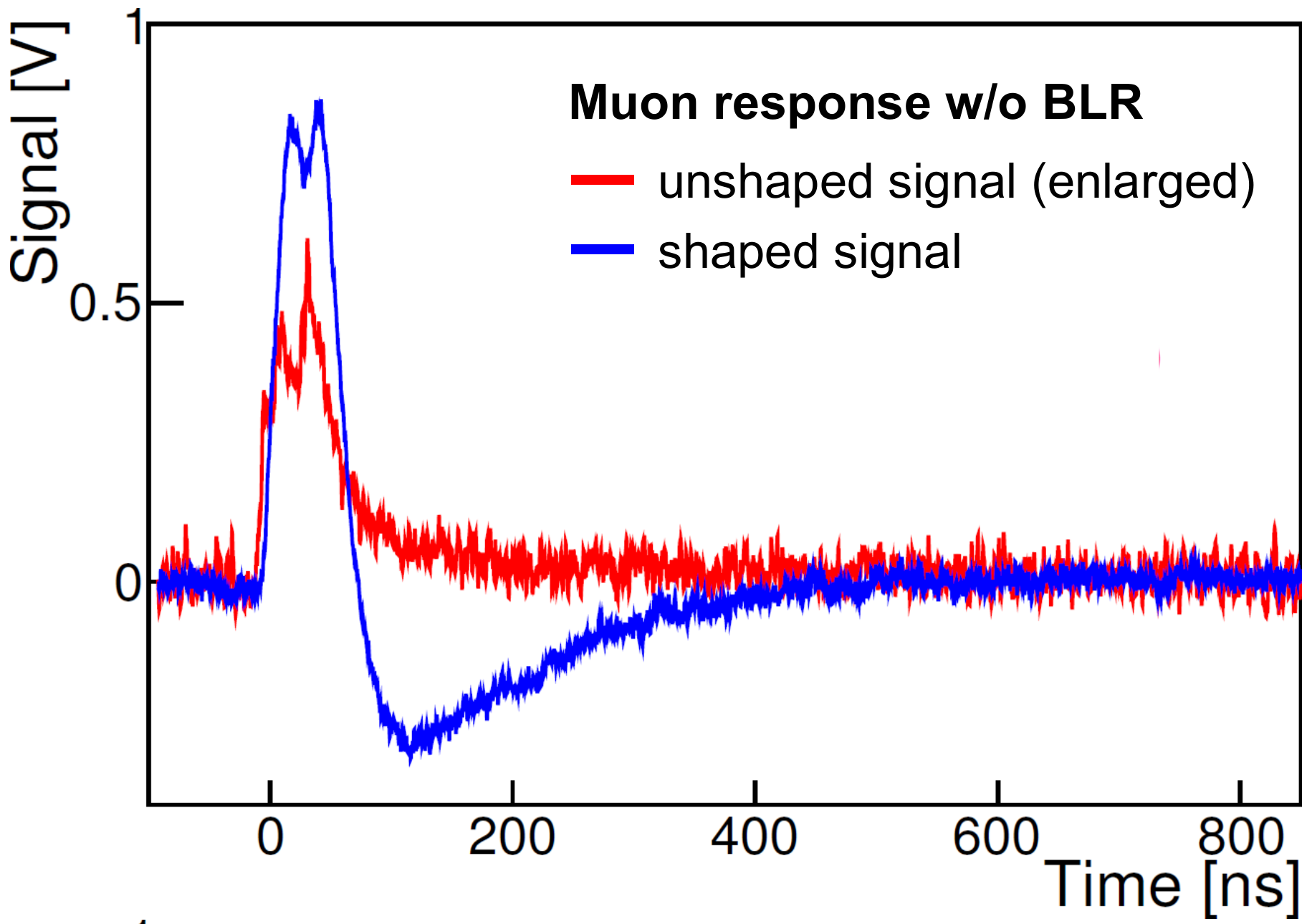}\\
\vspace{3mm}
\includegraphics[width=0.7\columnwidth]{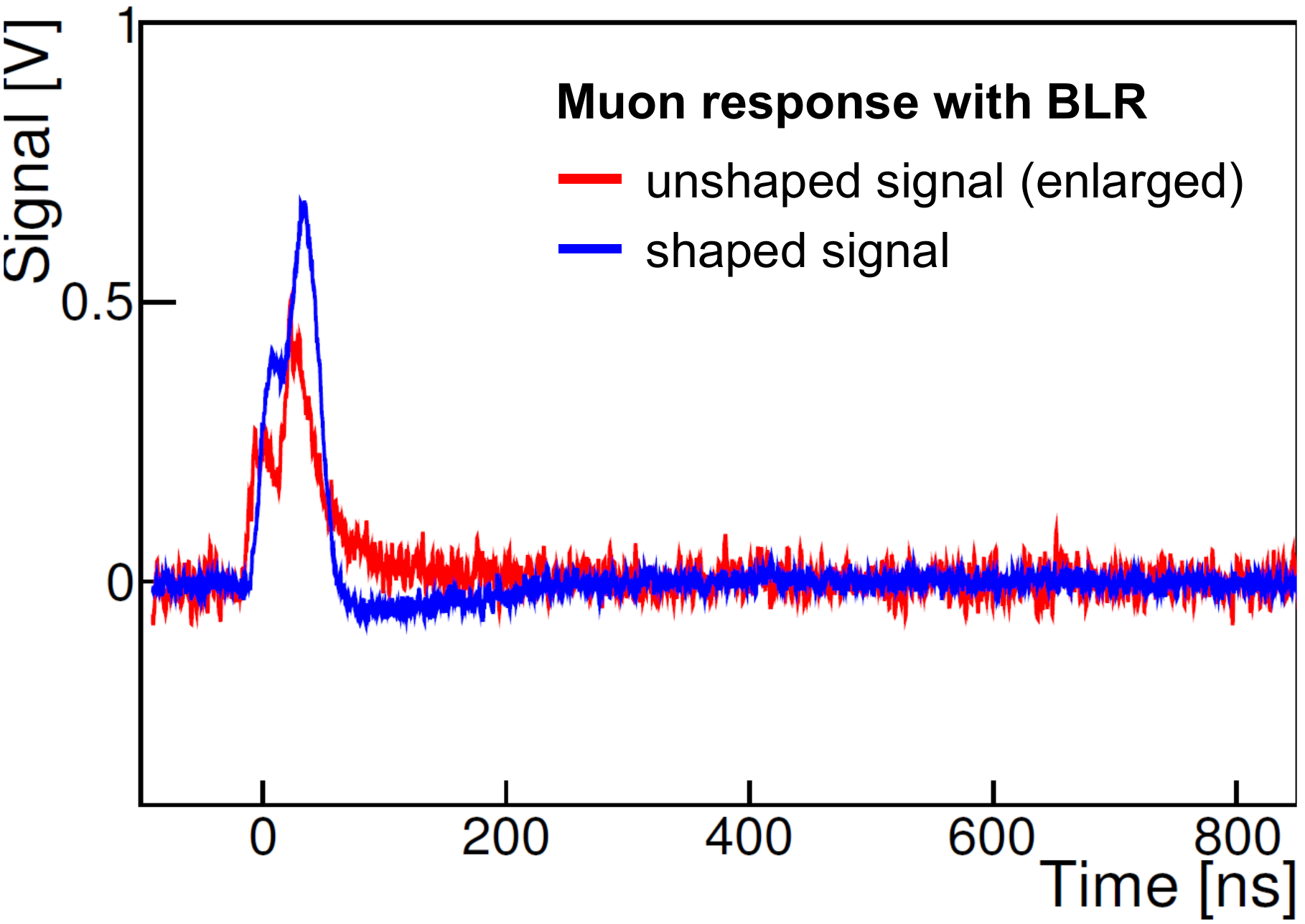}
\caption{Signals from an sMDT tube read out without (top) and with (bottom) baseline restoration (BLR)~\cite{cite:blr}.}
\label{fig:blr}
\end{figure}
An implementation of the BLR has been tested in a setup with an sMDT chamber at the GIF facility~\cite{cite:blr}. The discrete circuit consists of components similar to the existing ASD chip, but with an additional BLR stage between shaping and discrimination, and a slightly improved peaking time. In Figure~\ref{fig:blr}, the signal is shown after pre-amplification with a gain of $10^{5}$, and after the BLR stage. The BLR is activated applying a diode current $I_\mathrm{base}$. Comparing the shaped signal with and without activated BLR, it can be seen that the time to restore the baseline is reduced by a factor of two.
This demonstrates that with an improved read-out electronics, the shorter drift time for sMDT can be fully exploited.

\section{Chamber Design and Construction}
The chamber design and construction procedures have been optimized for mass production while providing a precision of better than 10 micrometers in the sense wire positions and the mechanical stability required to cover large areas. The design of the sMDT chambers profits from the long experience with the MDT chambers in ATLAS and provides even higher reliability. 

The tube design uses industrial components on one hand (e.g. standard Al tubes), and dedicated machined components that can be easily manufactured at high precision on the other. The central component is the wire locator in the sealing end, consisting of an inner part inside the tube, and an outer part which provides a reference surface. Both feature a very small tolerance, which allows to measure the sense wire position of the assembled tube with an accuracy of a few micrometer. 

\begin{figure}[tb]
\centering
\includegraphics[width=0.9\columnwidth]{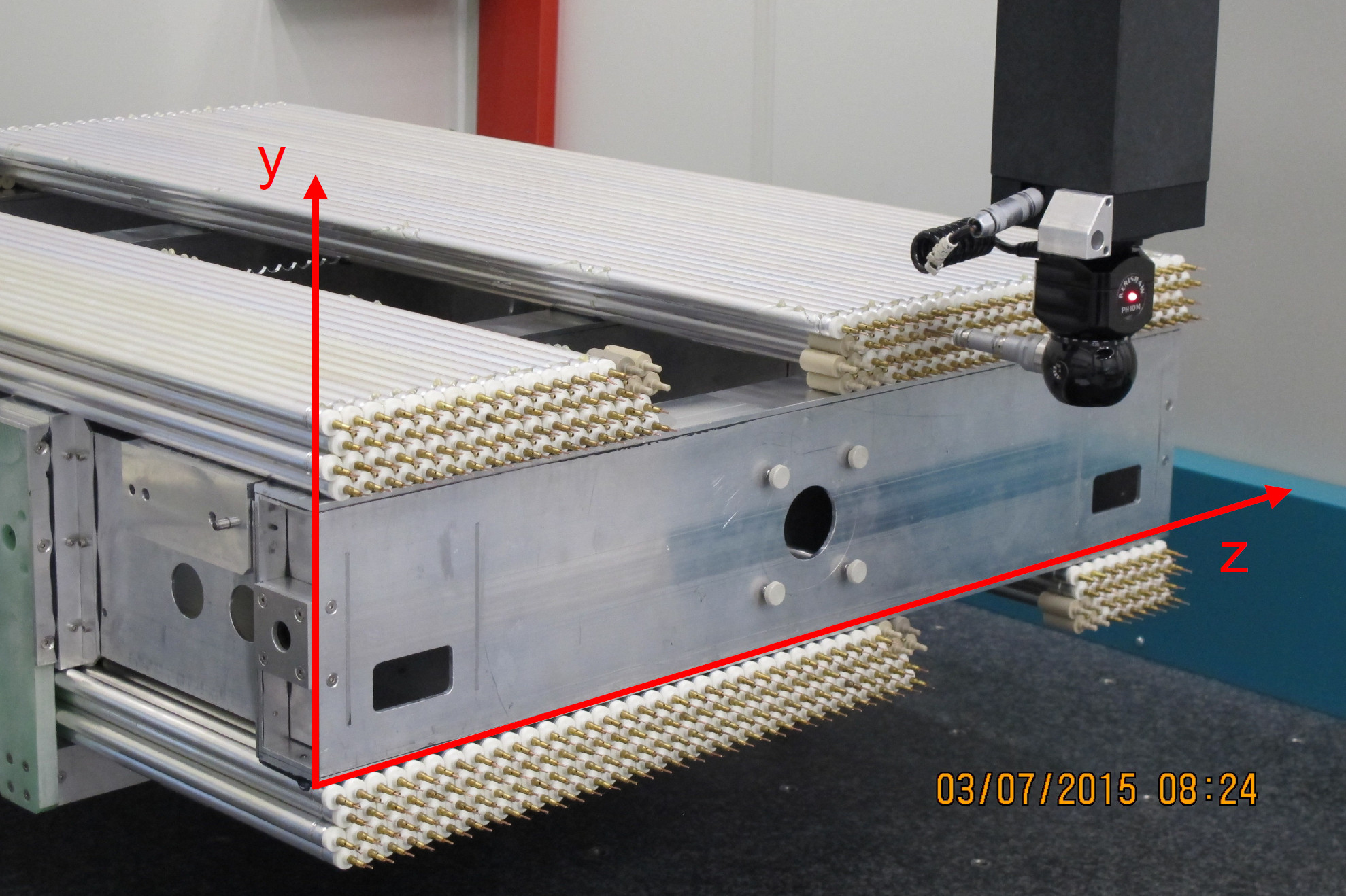}
\caption{Measurement of the sense wire position at a finished sMDT chamber using an automated mechanical sensor. The red arrows show the coordinate system.}
\label{fig:chamber}
\end{figure}
\begin{figure}[tb]
\centering
\includegraphics[width=0.9\columnwidth]{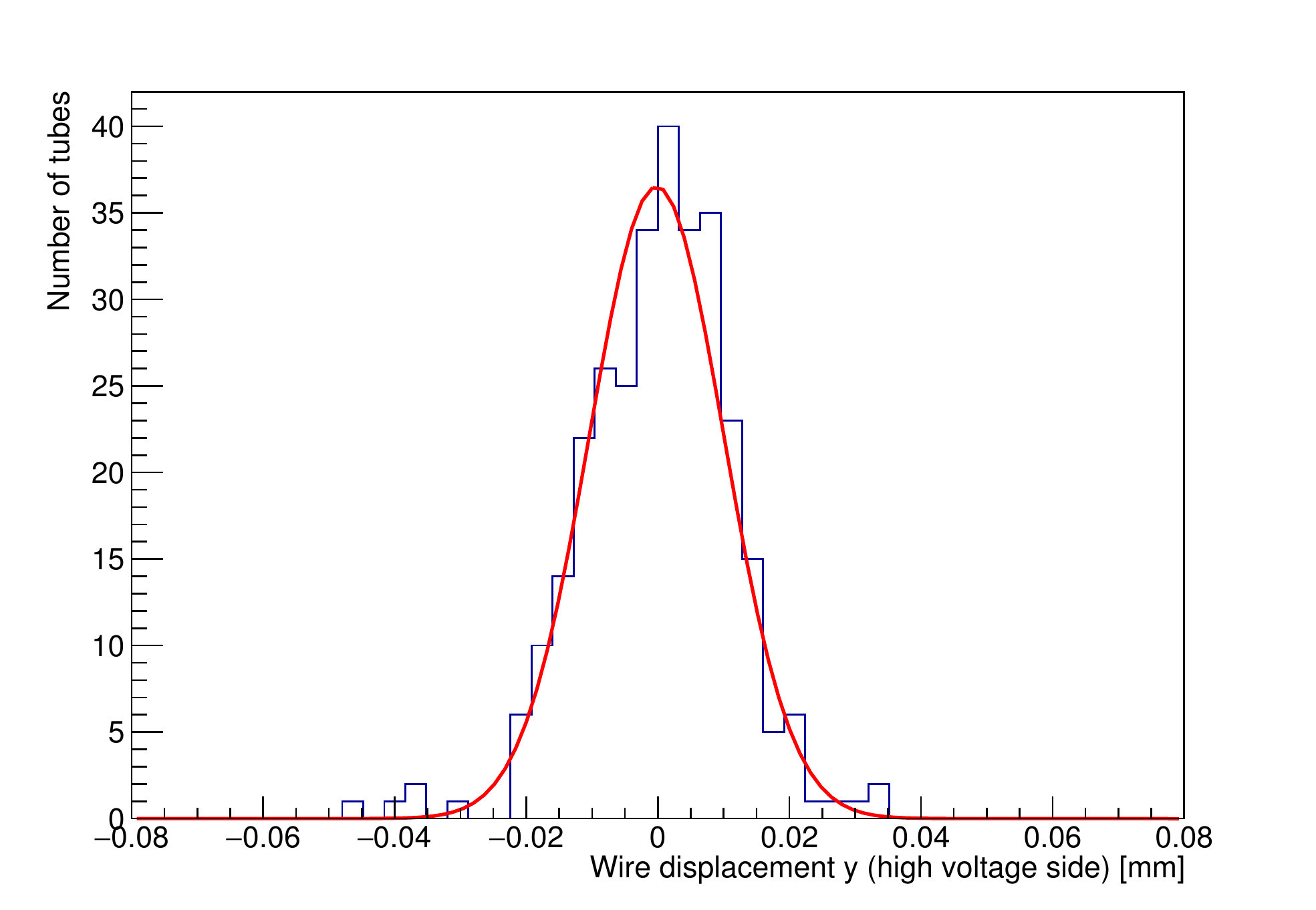}
\caption{Sense wire position in horizontal direction for a finished sMDT chamber.}
\label{fig:positioning}
\end{figure}
sMDT chambers are under construction for upgrades of the muon spectrometer of the ATLAS detector at the HL-LHC. The drift tube assembly is a semi-automated process with minimal manual intervention. Several sMDT chambers have already been produced and installed for the current LHC data taking. The mechanical precision of the sense wire positions has been measured using the external reference surface (see Fig.~\ref{fig:chamber}). The displacement is found to be $z_\mathrm{RMS}<10\,\upmu\mathrm{m}$ in horizontal and $y_\mathrm{RMS}<15\,\upmu\mathrm{m}$ in vertical direction. Figure~\ref{fig:positioning} shows the according measurement for the vertical displacement. The inherent mechanical precision allows for highly accurate monitoring of the absolute alignment of the chambers in the detector. 

\section{Summary}
Future colliders will put stringent requirements on muon detectors in terms of efficiency, spatial resolution, rate capabilities and radiation hardness. The sMDT chambers are a well-suited technology for large-area precision muon tracking at high background rates. Results from the GIF measurement show that sMDT provide both, the required spatial resolution and efficiency, as well as the rate capabilities and radiation hardness. The assembly process have been optimized for mass production and show an excellent mechanical precision. Hence, sMDT represent a cost-effective technology for large detector areas and and are a perfect candidate for future hadron colliders.

\ifCLASSOPTIONcaptionsoff
  \newpage
\fi

\end{document}